\documentclass[preprint,floatfix,showpacs,amsmath,amssymb,aip]{revtex4-2}
\usepackage[usenames,dvipsnames]{color}
\usepackage{graphicx,textcase} 
\usepackage{dcolumn,overpic} 
\usepackage{bm,color}
\usepackage[T1]{fontenc}
\usepackage{ae,aecompl}
\usepackage{chngcntr}
\usepackage{color,soul}
\usepackage[para]{threeparttable}
\usepackage{etoolbox}
\usepackage{lipsum}
\AtEndEnvironment{table*}{\vskip10pt}{}{}
\usepackage[bottom]{footmisc}
\usepackage{setspace}
\AtEndEnvironment{table*}{\vskip10pt}{}{}

\begin{document}
\title{Magnetic properties of the Fe$_5$SiB$_2$--Fe$_5$PB$_2$ system}

\author{Daniel Hedlund}
	\affiliation{Department of Engineering Sciences, Uppsala University, Box 534, 751 21 Uppsala, Sweden}
\author{Johan Cedervall}
	\affiliation{Department of Chemistry - \AA ngstr\"{o}m Laboratory, Uppsala University, Box 538, 751 21 	Uppsala, Sweden.}
\author{Alexander Edstr\"{o}m}
	\affiliation{Division of Materials Theory, Department of Physics and Astronomy, Uppsala University, 		Box 516, SE-751 20, Uppsala, Sweden}
	\affiliation{Department of Materials Science, ETH Zurich, Wolfgang-Pauli-Str. 27, 8093 Z\"{u}rich, Switzerland}
	\author{Miros\l{}aw Werwi\'nski}
	\affiliation{Institute of Molecular Physics, Polish Academy of Sciences, M. Smoluchowskiego 17, 60-179 Pozna\'{n}, Poland}
\author{Sofia Kontos}
	\affiliation{Department of Engineering Sciences, Uppsala University, Box 534, 751 21 Uppsala, Sweden}
\author{Olle Eriksson}
	\affiliation{Division of Materials Theory, Department of Physics and Astronomy, Uppsala University, 		Box 516, SE-751 20, Uppsala, Sweden}
	\affiliation{School of Science and Technology, \"Orebro University, SE-701 82 \"Orebro, Sweden}
\author{J\'an Rusz}
	\affiliation{Division of Materials Theory, Department of Physics and Astronomy, Uppsala University, 		Box 516, SE-751 20, Uppsala, Sweden}
\author{Peter Svedlindh}
	\affiliation{Department of Engineering Sciences, Uppsala University, Box 534, 751 21 Uppsala, Sweden}
\author{Martin Sahlberg}
	\affiliation{Department of Chemistry - \AA ngstr\"{o}m Laboratory, Uppsala University, Box 538, 751 21 	Uppsala, Sweden.}
\author{Klas Gunnarsson}
	\affiliation{Department of Engineering Sciences, Uppsala University, Box 534, 751 21 Uppsala, Sweden}

\date{\today}
\begin{abstract}
The magnetic properties of the compound Fe$_5$Si$_{1-x}$P$_{x}$B$_2$ have been studied, with a focus on the Curie temperature $T_\textrm{C}$, saturation magnetization $M_\textrm{S}$, and magnetocrystalline anisotropy. Field and temperature dependent magnetization measurements were used to determine $T_\textrm{C}\left(x\right)$ and $M_\textrm{S}\left(x\right)$. The saturation magnetization at 10 K (300 K) is found to monotonically decrease from $1.11~\mathrm{MA/m}$ ($1.03~\mathrm{MA/m}$) to $0.97~\mathrm{MA/m}$ ($0.87~\mathrm{MA/m}$), as $x$ increases from zero to one. The Curie temperature is determined to be 810 K and 615 K in Fe$_5$SiB$_2$ and Fe$_5$PB$_2$, respectively. The highest $T_\textrm{C}$ is observed for $x=0.1$, while it decreases monotonically for larger $x$. The Curie temperatures have also been theoretically determined to be 700 K and 660 K for Fe$_5$SiB$_2$ and Fe$_5$PB$_2$, respectively, using a combination of density functional theory and Monte Carlo simulations. The magnitude of the effective magnetocrystalline anisotropy was extracted using the law of approach to saturation, revealing an increase with increasing phosphorus concentration. Low--field magnetization vs. temperature results for $x = 0, 0.1, 0.2$ indicate that there is a transition from easy--axis to easy--plane anisotropy with decreasing temperature.
\end{abstract}
\maketitle

\section{Introduction}
Magnetically ordered materials are key to the modern world. Some examples include magnetic materials used in electric generators, electric motors, hard drives and medical devices such as magnetic resonance imaging. Exploration of the known and unknown magnetic materials is essential for further development of magnetic technology, as well as for the enhanced understanding of magnetic phenomena.

One system with potential for applications is the Fe$_5$SiB$_2$--Fe$_5$PB$_2$ system, which has attracted interest because of a Curie temperature $T_{\mathrm{C}}$ higher than 760 K (Fe$_5$SiB$_2$), a saturation magnetization $M_{\mathrm{S}}$ larger than 1 MA/m and a magnetocrystalline anisotropy (MAE) \textgreater 0.30 MJ/m$^3$ at room temperature.\cite{Cedervall2016,McGuire2015,PhysRevB.93.174412,lamichhane_study_2016} Fe$_5$SiB$_2$ was first synthesized in 1959 by Aronsson and Lundgren.\cite{aronsson1959} Using M\"{o}ssbauer spectroscopy W\"{a}ppling \emph{et al.}\cite{wappling1976} determined that Fe$_5$SiB$_2$ is ferromagnetic up to the reported $T_{\mathrm{C}}$ of 784 K and that the magnetization is parallel to the c-axis at high temperatures. Other studies include experiments on single crystalline Fe$_5$PB$_2$ by Lamichhane \emph{et al.}\cite{lamichhane_study_2016} who reported a $T_\mathrm{C}$ of 655 K, an anisotropy constant $K_1$ of 0.38 MJ/m$^3$ and $M_\textrm{S}=0.92$ MA/m at 300 K.\cite{lamichhane_study_2016} McGuire \emph{et al.}\cite{McGuire2015} studied the Fe$_5$SiB$_2$--Fe$_5$PB$_2$ system and they reported $T_\textrm{C}$ \textgreater  760 K (maximum temperature in their experiments) for Fe$_5$SiB$_2$ and 640 K for Fe$_5$PB$_2$. McGuire \emph{et al.} also studied the effect of substitution of Fe for Co and Mn in Fe$_5$PB$_2$ and found that $T_\mathrm{C}$ decreased down to 515 K for Fe$_4$CoPB$_2$, while $T_\mathrm{C}$ increased to 650 K for Fe$_4$MnPB$_2$. 

\begin{figure}[p]
	\centering
\includegraphics[width = 0.60\textwidth]{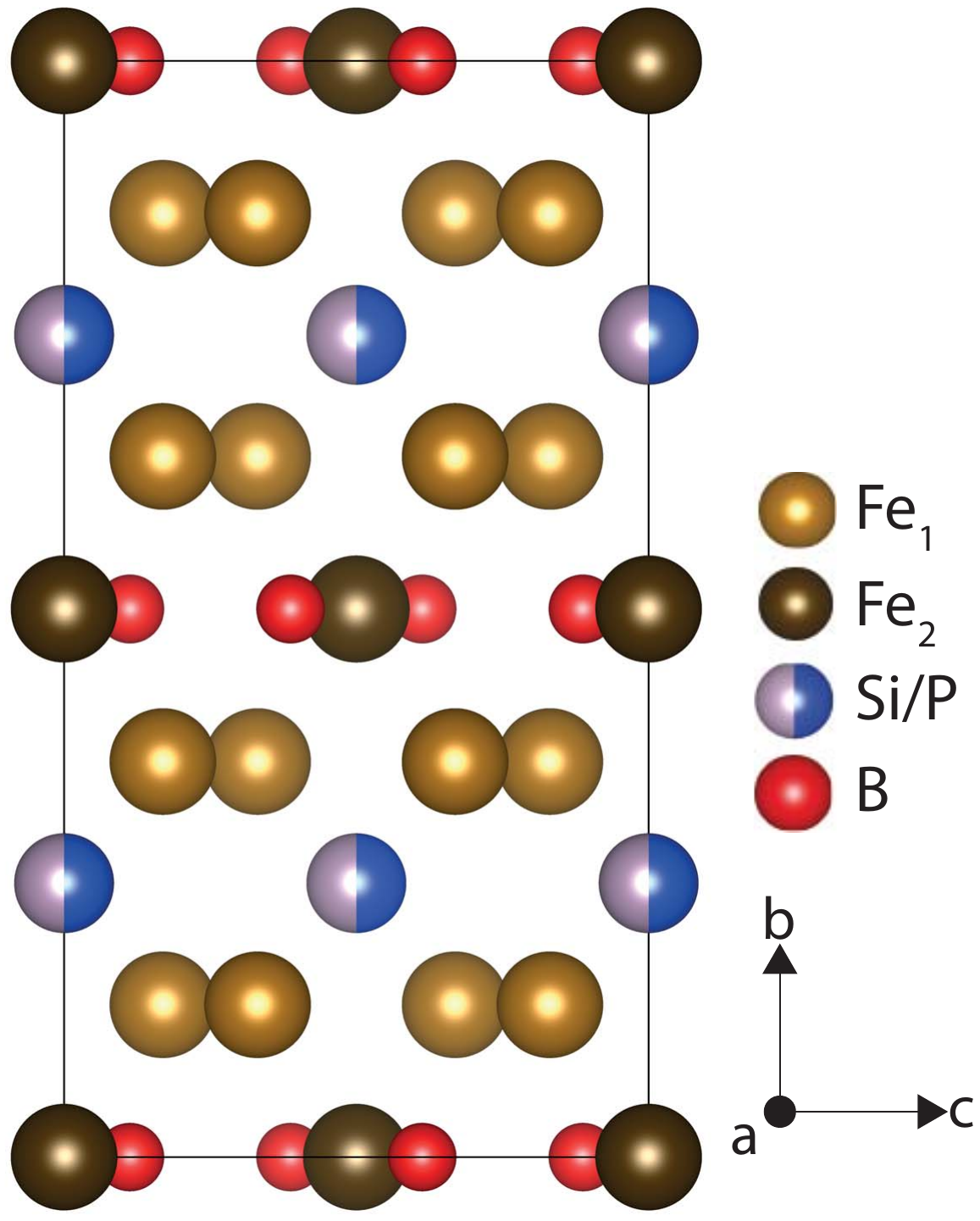}
\caption{Crystal structure of Fe$_5$Si$_{1-x}$P$_x$B$_2$.}
\label{fig:structure}
\end{figure}

Fe$_5$SiB$_2$ is part of a larger system of borides, M${_5}$XB${_2}$, where $M$ is a $d$--metal such as Fe, Co or W, and $X$ is a P--block element such as P or Si. The crystal structure is shown in Fig. \ref{fig:structure}, where the two Fe positions are marked, as well as the B position and the crystallographic site that contains the Si and P atoms in the compound. 
Using density functional theory with the virtual crystal approximation (VCA) Werwi\'nski \emph{et al.} \cite{PhysRevB.93.174412} reported that the MAE varied from -0.3 MJ/m$^3$ to 0.3 MJ/m$^3$  as Si was  substituted with an increasing amount of P. In this work it was also suggested that Fe$_5$S$_{1-x}$P$_x$B$_2$ compounds with sulfur can have an MAE as high as 0.8 MJ/m$^{3}$.

In this study, the Fe$_5$Si$_{1-x}$P$_x$B$_2$ system with $0 \leq x \leq 1$ has been experimentally investigated using temperature and magnetic field dependent magnetization measurements. Results from temperature dependent magnetometry between $T=10$ K and $T=950$ K have been used to determine the variation of $T_\textrm{C}$ with $x$. Field dependent magnetization measurements performed at $T=10$ K and $T=300$ K were used to determine the variation of $M_\textrm{S}$ with changing phosphorus concentration. The law of approach to magnetic saturation was used to determine the variation of the effective magnetocrystalline anisotropy with $x$. Phase analysis has been performed with X--ray diffraction (XRD) and thermogravimetric analysis (TGA).  Density functional theory (DFT) calculations have been performed using the spin polarized relativistic Korringa-Kohn-Rostoker (KKR) method\cite{sprkkr, Ebert2011} and the generalized gradient approximation.\cite{PhysRevLett.77.3865} The Curie temperatures have been evaluated using the Heisenberg exchange interaction parameters extracted from the DFT results and using these as input for classical Monte Carlo (MC) simulations.

\section{Experiment}
\subsection{Sample preparation and material characterization}
Master compounds of Fe$_5$SiB$_2$ and Fe$_5$PB$_2$ were prepared from stoichiometric amounts of the elements iron (Leico Industries, purity 99.995\%. Surface oxides were reduced in H$_{2}$ gas.), silicon (Highways International, purity 99.999\%), phosphorus (Cerac, purity 99.999\%) and boron (Wacher-Chemie, purity 99.995\. Fe$_5$SiB$_2$ was prepared by conventional arc melting while drop synthesis \cite{Carlsson1973} was used for Fe$_5$PB$_2$, where phosphorus was dropped in to a melt of iron and boron. Samples of intermediate compositions in the series Fe$_5$Si$_{1-x}$P$_x$B$_2$ were made by mixing appropriate amounts of the master alloys. All samples were crushed, pressed into pellets and heat treated in evacuated silica tubes at 1273 K for 33 days followed by quenching in cold water. The pure Fe$_5$SiB$_2$ sample is the same as used in Ref.~[\onlinecite{Cedervall2016}].

Phase analysis and crystal structure determinations were performed using a Bruker D8 diffractometer equipped with a LynxEye position sensitive detector  (PSD, 4$^{\circ}$ opening) using CuK$\alpha_1$ radiation ($\lambda$ = 1.540598 \AA). All measurements were performed at 298 K in a 2$\theta$ range of 20-90$^{\circ}$. The crystal structures were studied in detail with refinements according to the Rietveld method \cite{Rietveld1969} implemented in the software FULLPROF.\cite{Rodriguez-Carvajal1993}
In the refinements 18 parameters were varied: 5 atomic coordinates, 4 atomic occupancies, isotropic temperature factor and 2 unit cell parameters as well as the zero point, background, scale factor, peak shape and 3 half width parameters. The unit cell parameters of the main phase were precisely studied using least square refinements of the peak positions using the software UNITCELL.\cite{Holland1997}

To evaluate the melting point of the samples, differential thermal analysis (DTA) was performed using a Netzsch STA 409 PC Luxx TG-DTA/DSC instrument. All measured samples were analysed under flowing Ar atmosphere and with Ar also used as a purge gas in the temperature range 298 to 1668 K with a heating rate of 10 K/min.

\subsection{Magnetization measurements}
All samples were studied using vibrating sample magnetometry.
Field ($H$) and temperature ($T$) dependent magnetization ($M$) measurements were performed using two different vibrating sample magnetometers (VSMs). Temperature dependent values of $M$ between 10 K $\leq T \leq$ 300 K were measured using a Quantum Design PPMS 6000 VSM, while measurements between 300 K $\leq T \leq$ 950 K were performed using a LakeShore 7404 VSM. Samples measured at low temperature were immobilized in gelatin capsules using a weakly paramagnetic varnish that accounts for less than 0.01\% of the measured magnetic moment. The samples used for $T$ > 300 K  were folded in 0.05 mm Cu--foil during measurements. Full hysteresis curves were measured at a temperature of 300 K using the LakeShore VSM with fields ranging from 1.4 MA/m to -1.4 MA/m. None of the samples exhibited a hysteric behavior and therefore only positive magnetic fields were applied using the PPMS VSM at $T=10$ K and $T=300$ K. The magnetization in SI units was calculated from the magnetic moment by using the sample weight and the density obtained from XRD measurements at 298 K. High temperature measurements were performed to extract the $T_\mathrm{C}$ whereas low temperature were made to reveal the magnetostructural transitions previously reported for this system.\cite{Cedervall2016} Using the law of approach to saturation \cite{chikazumi_physics_1997} an effective anisotropic constant $\left| {K}_{\mathrm{eff}} \right|$ was determined using the same method as in Ref. [\onlinecite{Cedervall2016}].

\subsection{Details of calculations}
The DFT calculations were carried out in the generalized gradient approximation\cite{PhysRevLett.77.3865}  using the spin polarized relativistic KKR method\cite{sprkkr, Ebert2011} in scalar relativistic mode.
Previous work has shown that the atomic sphere approximation (ASA) is not accurate in describing the relevant system,\cite{PhysRevB.93.174412} resulting in e.g. incorrect magnetic moments in comparison to full potential (FP) calculations. In other systems, such as (Fe$_{1-x}$Co$_{x}$)$_2$B, ASA has been shown to provide an unsatisfactory description of the band structure and certain delicate magnetic properties including the magnetocrystalline anisotropy.\cite{Edstrom2015} Since not even the magnetic moments are correctly described by the ASA for the system studied here, the calculations have been performed in the more accurate FP mode. Previously reported~\cite{PhysRevB.93.174412} computationally optimized lattice parameters and atomic positions were used. 
For the self-consistent calculations 40 energy points on a semi-circular path in the complex energy plane and a discretization over approximately 1000 $k$-points in the Brillouin zone were considered.
$J_{ij}$, the exchange interactions, appearing in the Heisenberg Hamiltonian, were evaluated by the method of Liechtenstein \emph{et al.},\cite{Liechtenstein1984, Liechtenstein1986} with respect to a ferromagnetic reference state.

\section{Results and discussion}

\subsection{Phase analysis}
\begin{figure} 
	\centering
\includegraphics[width = 1.00\textwidth]{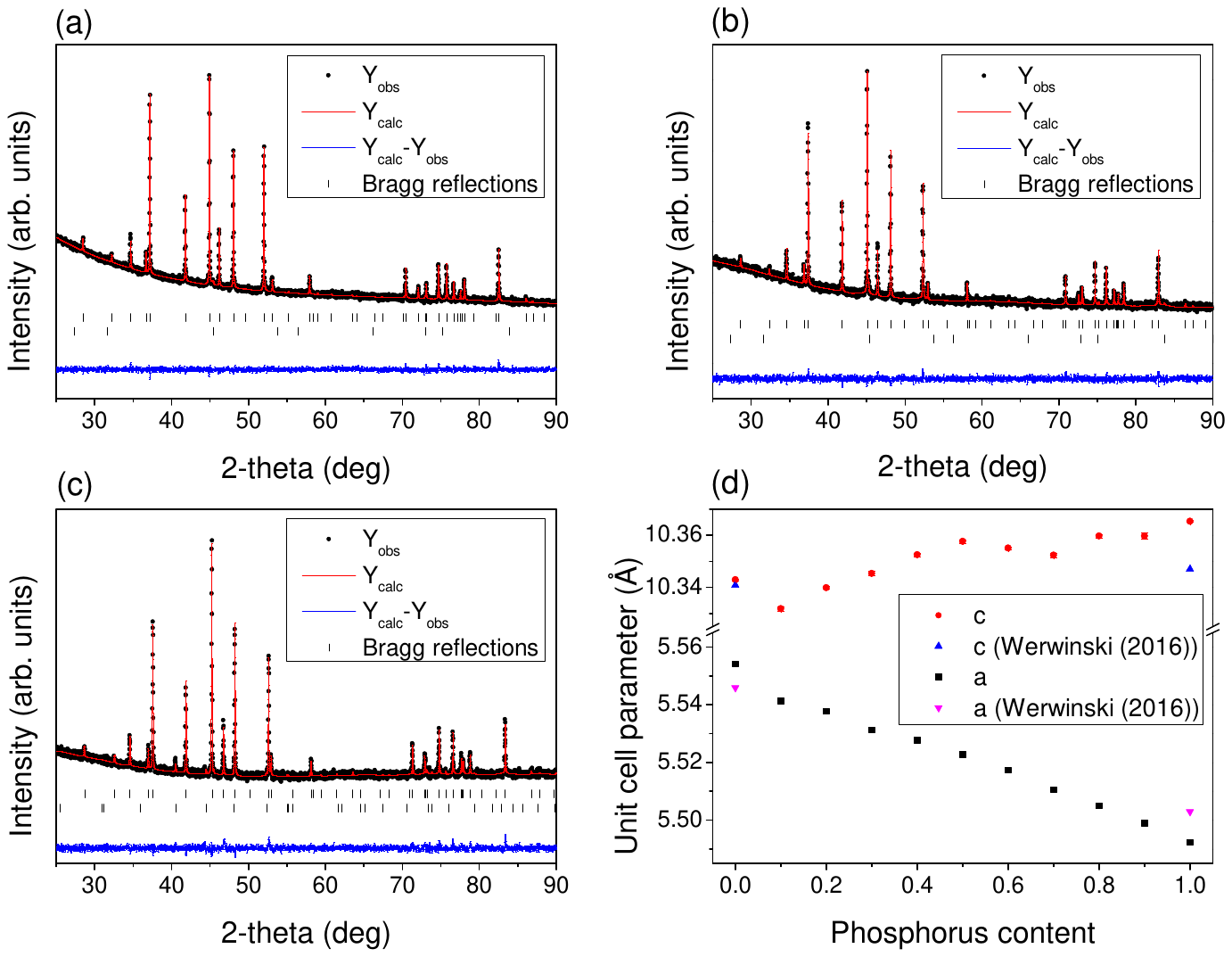}
\caption{Structure refinements performed on the samples (a) Fe$_5$SiB$_2$, (b) Fe$_5$Si$_{0.5}$P$_{0.5}$B$_2$ and (c) Fe$_5$PB$_2$. (d) shows the refined unit cell parameters as a function of phosphorous content. The secondary phases are Fe$_3$Si in (a) and (b) and Fe$_2$P in (c).}
\label{fig:XRD}
\end{figure}

\begin{figure}[b]
	\centering
 \includegraphics[width = 1.00\textwidth]{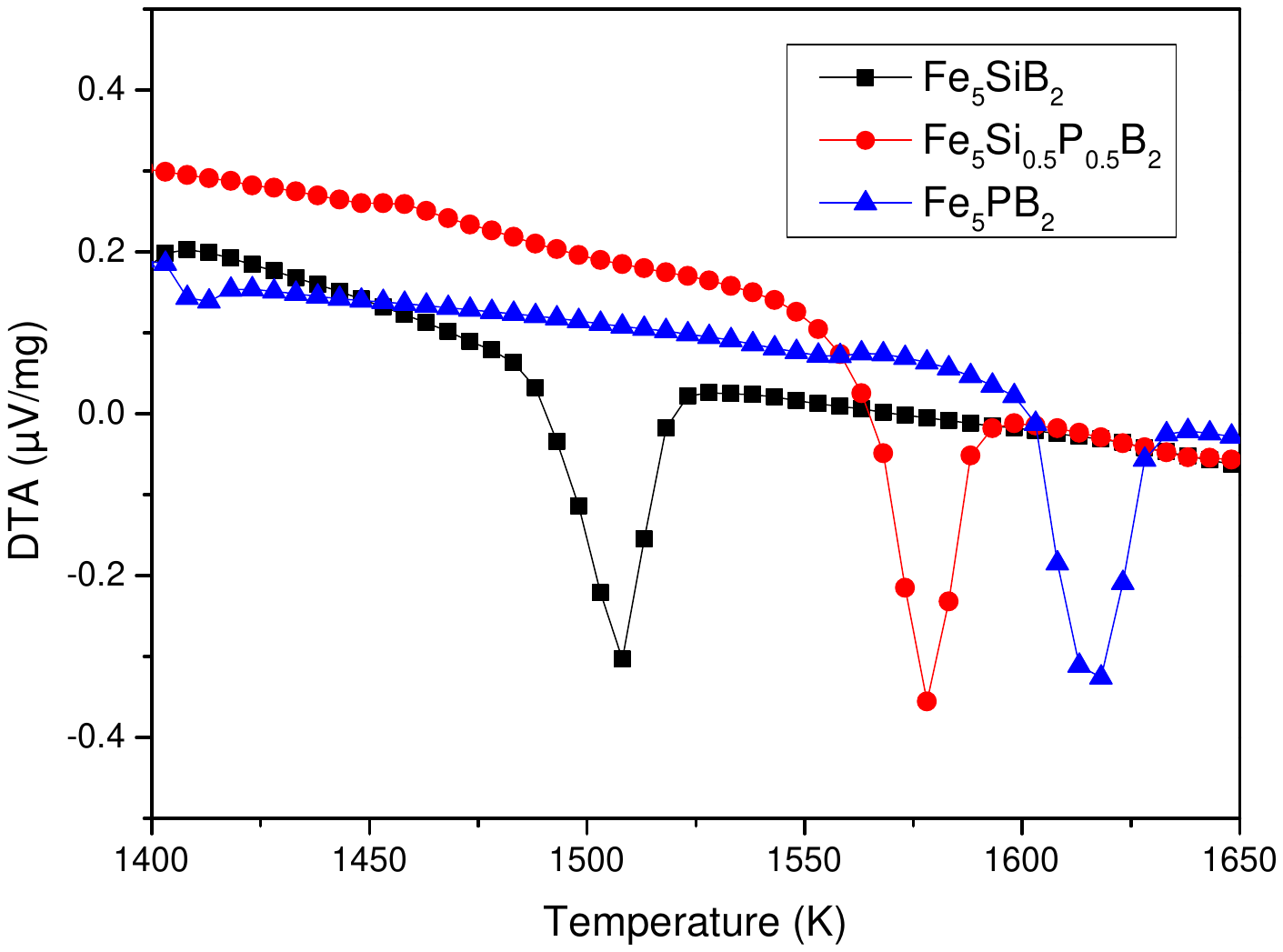}
 \caption{DTA scans used to extract the melting temperatures for the samples Fe$_5$SiB$_2$, Fe$_5$Si$_{0.5}$P$_{0.5}$B$_2$ and Fe$_5$PB$_2$.}
 \label{fig:DTA}
\end{figure}
The X-ray diffraction (XRD) studies confirm that the master compounds crystallize in the $I$4/$mcm$ space group  and the crystallographic main phase is the same for all the samples in the series.
The crystal structure analysis is shown in Fig. \ref{fig:XRD} for the master alloys Fe$_5$SiB$_2$ (a) and Fe$_5$PB$_2$ (c) as well as for the mixed compound Fe$_5$Si$\rm _{0.5}$P$\rm _{0.5}$B$_2$ (b).
The crystal structure analysis has, however, shown that all samples have secondary phases (\textless 3\%). 
For Fe$_5$SiB$_2$ and Fe$_5$Si$_{0.5}$P$_{0.5}$B$_2$ it consists of Fe$_3$Si while for Fe$_5$PB$_2$ the secondary phase is Fe$_2$P.
The unit cells of all samples have been refined and show that the $a$ lattice parameter decreases linearly while the $c$ lattice parameter increases slightly when going towards higher phosphorous concentrations, see Fig. \ref{fig:XRD} (d).
For the Fe$_5$SiB$_2$ sample the unit cell parameters are $a=5.5541(1)~\text{\AA}$ and $c=10.3429(2)~\text{\AA}$.\cite{Cedervall2016}
These values change to $a=5.4923(1)~\text{\AA{}}$ and $c=10.3654(4)~\text{\AA{}}$ when the silicon site is fully substituted with phosphorous. 
This trend was also previously calculated by Werwi\'nski \emph{et al.}\cite{PhysRevB.93.174412} 
The fact that phosphorous is replacing silicon in the structure is further supported by the melting temperatures extracted from the DTA results presented in Fig. \ref{fig:DTA}.
It becomes clear since the lowest melting temperature is observed for Fe$_5$SiB$_2$ (1491 K) and the highest for Fe$_5$PB$_2$ (1603 K), with the intermediate sample (Fe$_5$Si$\rm _{0.5}$P$\rm _{0.5}$B$_2$) exhibiting a melting temperature in between (1566 K).
 
\subsection{Experimental and calculated magnetic results}
\begin{figure} 
\centering
\includegraphics[width = 0.80\textwidth]{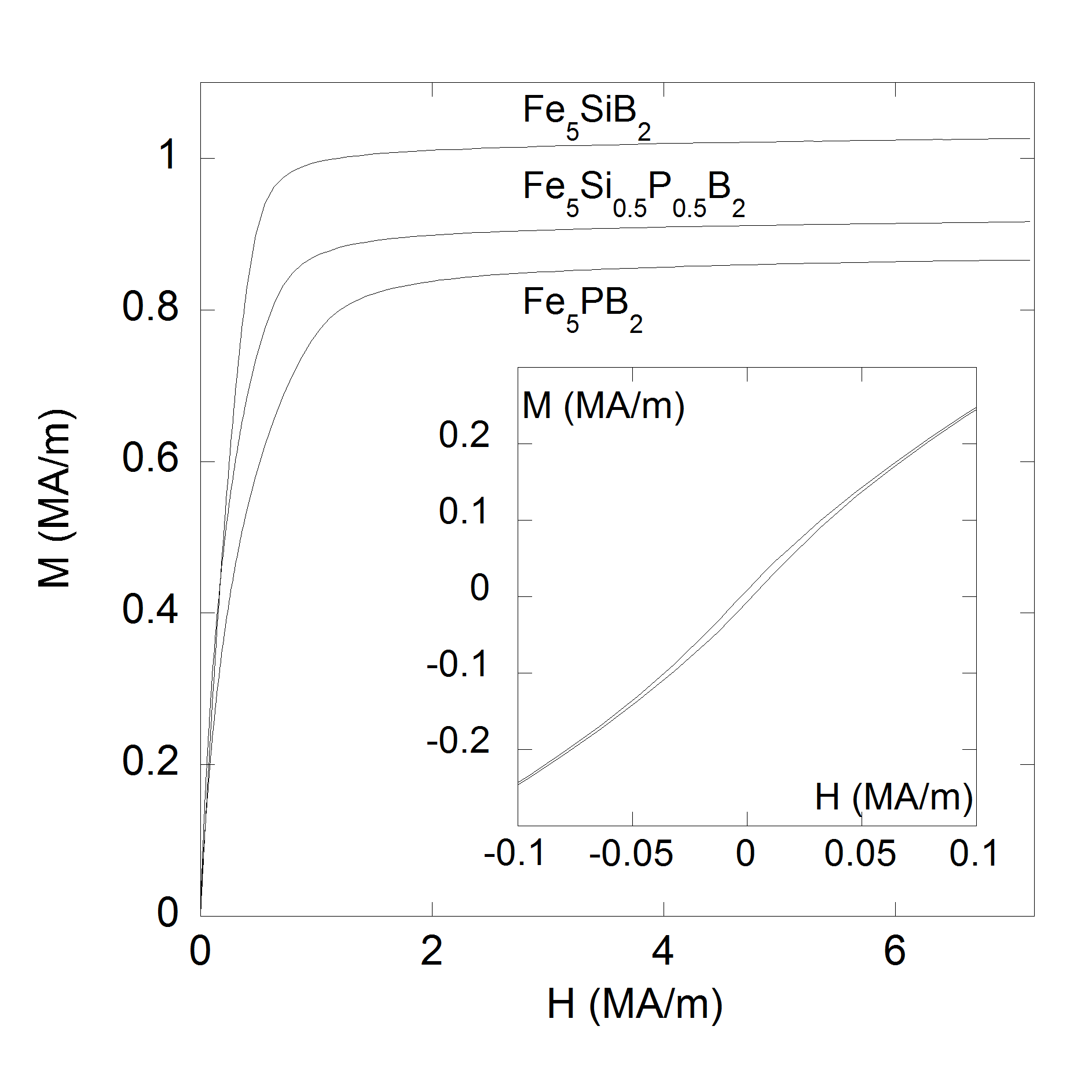}
\caption{ $M$ vs. $H$ at 300 K from top to bottom for Fe$_5$SiB$_2$, Fe$_5$Si$_{0.5}$P$_{0.5}$B$_2$ and 
	Fe$_5$PB$_2$.
	Insert shows the $M-H$--curve for Fe$_5$Si$_{0.5}$P$_{0.5}$B$_2$ at weak fields.}
\label{fig:mh}
\end{figure}

In Fig. \ref{fig:mh}, magnetization vs. field curves for Fe$_5$SiB$_2$, Fe$_5$Si$_{0.5}$P$_{0.5}$B$_2$ and Fe$_5$SiB$_2$ at 300 K are shown.
All samples exhibit close to zero coercivity and remanence, which is illustrated by the insert in Fig. \ref{fig:mh} for Fe$_5$Si$_{0.5}$P$_{0.5}$B$_2$. The highest saturation magnetization is observed for Fe$_5$SiB$_2$ ($M_{\mathrm{S}}=1.03$ MA/m). In a previous publication by Cedervall \emph{et al.}\cite{Cedervall2016} it was reported that $M_{\mathrm{S}}=1.02$ MA/m for Fe$_5$SiB$_2$ at 300 K, in good agreement with the present results. The present results can also be compared to those of McGuire \emph{et al.},\cite{McGuire2015} who reported a $M_{\mathrm{S}}$ of 0.96 MA/m (300 K) for Fe$_5$SiB$_2$. 
However, the Fe$_5$SiB$_2$ sample used in the present study contains  \textless3\% of the impurity phase Fe$_3$Si, a ferromagnetic compound with a reported $T_\mathrm{C}$ of about 840 K \cite{shinjo_magnetic_1963,liu2011}, which makes a direct comparison problematic. 
McGuire \emph{et al.}\cite{McGuire2015} found that their Fe$_5$SiB$_2$ sample contained 3\% Fe$_{4.7}$Si$_2$B. The values reported here for $M_{\mathrm{S}}$ are 6\% larger compared to those reported by McGuire \emph{et al}.\cite{McGuire2015} Regarding the  Fe$_{4.7}$Si$_2$B impurity phase, no reported data of its magnetization exists. 

In a recent work studying a Fe$_5$SiB$_2$ sample with isotope pure $^{11}$B,\cite{Cedervall2016} the $M_{\textrm{S}}$ was reported to be 0.92 MA/m at 300 K not taking into account the 5\% impurity of Fe$_{4.7}$Si$_2$B. 
Assuming  Fe$_{4.7}$Si$_2$B is non--magnetic and that different boron isotopes do not affect the magnetic moment then $M_{\mathrm{S}}$ would be 0.97 MA/m for Fe$_5$SiB$_2$. 
The $M_{\mathrm{S}}$ value reported by McGuire \emph{et al.} would be 1.08 MA/m at 300 K using the same reasoning and the reported value in the present work is 1.05 MA/m at 300 K. 
Since the difference between $M_{\mathrm{S}}$ using corrected and uncorrected moments are small no corrections were made for the other samples in the present work.

\begin{figure} 
	\centering
 \includegraphics[width=0.80\textwidth]{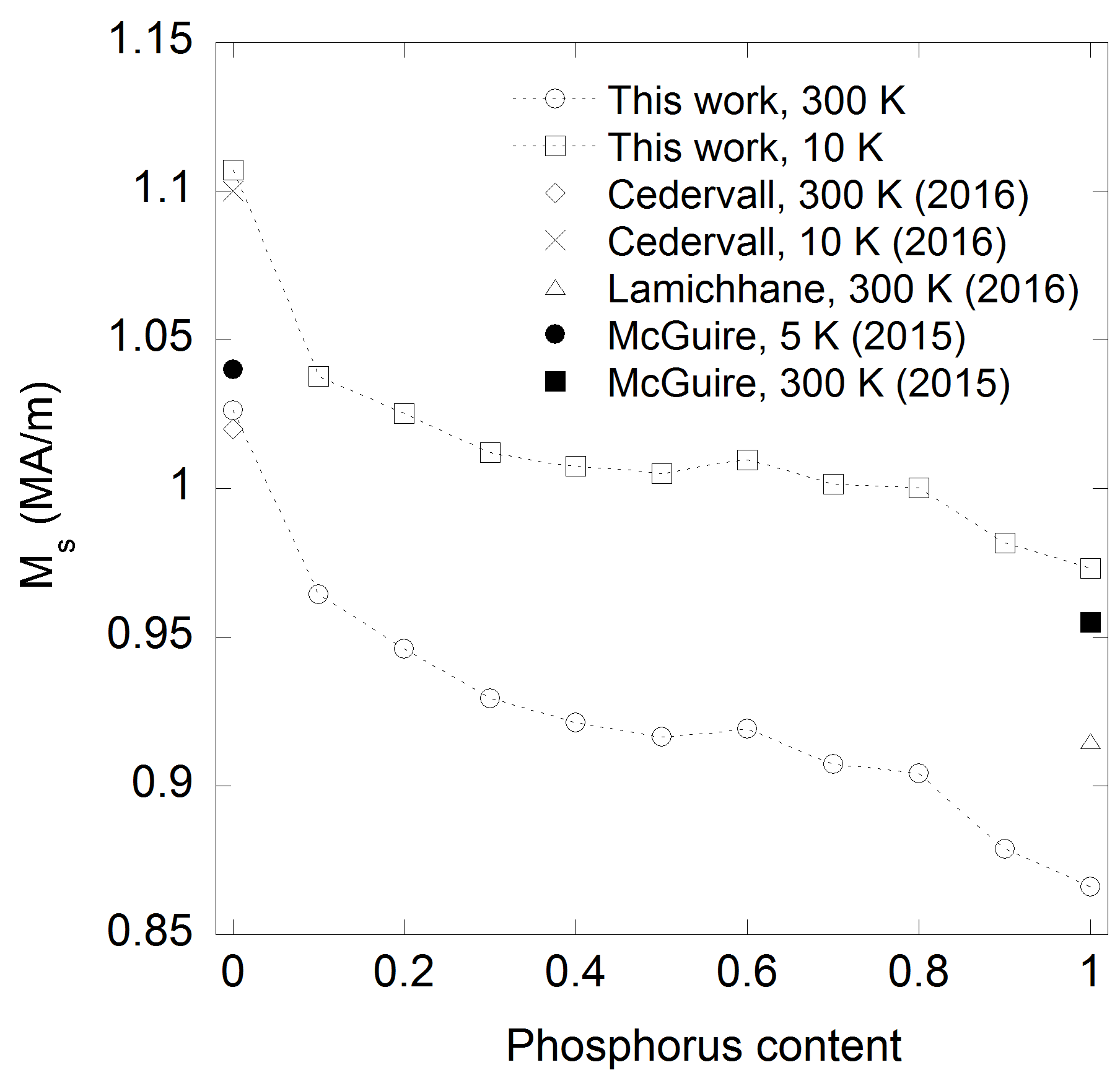}
 \caption{Experimental values for M$_{\mathrm{S}}$ at 300 K and 10 K together
 with previously reported values.\cite{Cedervall2016} Full substitution of Si for P decreases $M_{\mathrm{S}}$ from 1.03 MA/m (Fe$_5$SiB$_2$) to 0.87 MA/m (Fe$_5$PB$_2$) at 300 K. The dashed line is added as a guideline for the eye. }
 \label{fig:msat}
\end{figure}

Fig. \ref{fig:msat} displays the $M_{\mathrm{S}}$ values from this study together with previously reported values for the Fe$_5$SiB$_2$--Fe$_5$PB$_2$ system.
Previous studies have focused on Fe$_5$SiB$_2$ and Fe$_5$PB$_2$, with no compositions in between.
Taking all previous measurements into account, considering the impurity contents and the new results presented here it can be concluded that the saturation magnetization monotonously decreases with increasing P content (cf. Fig. \ref{fig:mh}). It should be pointed out that this decrease of the saturation magnetization with increasing phosphorus concentration does not comply with results from density functional theory calculations,\cite{PhysRevB.93.174412} where instead the results indicated a weak increase of the magnetization with phosphorus concentration (from 1.07 MA/m to 1.08 MA/m as $x$ increased from 0 to 1).

Regarding Fe$_5$PB$_2$, $M_{\mathrm{S}}$ is estimated to 0.87 MA/m and 0.97 MA/m at 300 K and 10 K, respectively. According to the XRD results in Fig. \ref{fig:XRD}, there is a secondary phase of Fe$_2$P present and the $M_{\mathrm{S}}$ values are therefore expected to be somewhat underestimated. The present results can be compared to those obtained from single crystal data of Fe$_5$PB$_2$ by Lamichhane \emph{et al.}\cite{lamichhane_study_2016}  who reported a $M_{\mathrm{S}}$ of 0.92 MA/m at 300 K and 1.03 MA/m at 5 K. 
The larger values of $M_{\mathrm{S}}$ obtained by Lamichhane \emph{et al.}\cite{lamichhane_study_2016} can be explained by single crystals being free from impurity phases. McGuire \emph{et al.}\cite{McGuire2015} on the other hand reported a $M_{\mathrm{S}}$ of 0.87 MA/m at 300 K and 0.96 MA/m at 5 K with 6\% Fe$_2$P impurity phase.
Substituting Si with P yields a reduction of $M_{\mathrm{S}}$ from 1.03 MA/m (Fe$_5$SiB$_2$) to 0.87 MA/m (Fe$_5$PB$_2$) at 300 K.
Since neither Si nor P are magnetic elements, the reduction of $M_{\mathrm{S}}$ should mainly originate from a change in unit cell volume, or from the magnitude of the magnetic moments on the Fe atoms. From XRD data the unit cell volume for Fe$_5$SiB$_2$ is 319.06 \AA$^3$  and 312.67 \AA$^3$ for Fe$_5$PB$_2$, indicating that the moments on the Fe atoms are larger in Fe$_5$SiB$_2$ as compared to Fe$_5$PB$_2$. 

\begin{figure} 
	\centering
 \includegraphics[width=0.80\textwidth]{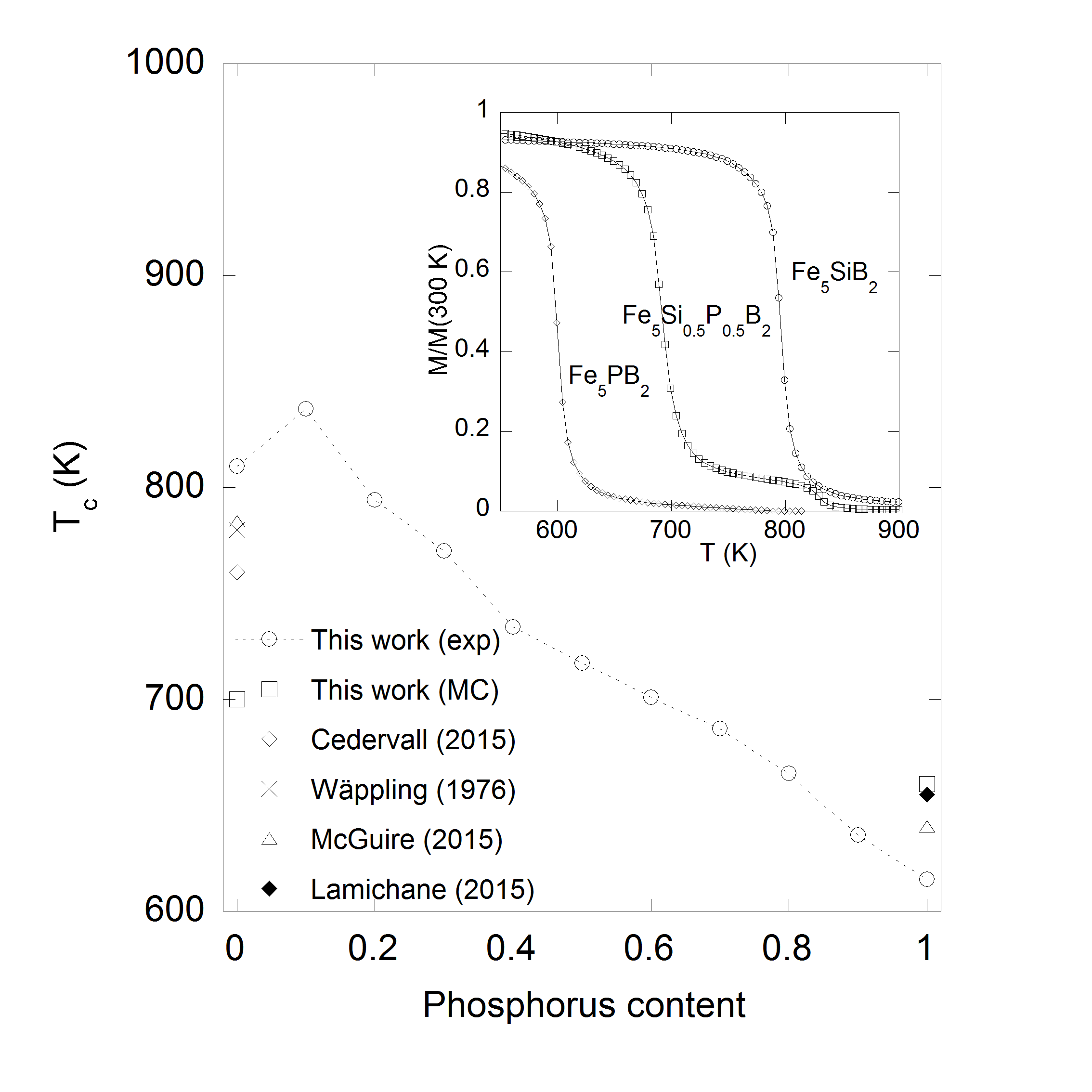}
 \caption{Experimental values of $T_{\mathrm{C}}$ as a function of P concentration together with values from Monte Carlo (MC) simulations. Insert shows $M$-$T$--curves for Fe$_5$SiB$_2$, Fe$_5$Si$_{0.5}$P$_ {0.5}$B$_2$ and Fe$_5$PB$_2$.} 
 \label{fig:tt}
\end{figure}

In Fig. \ref{fig:tt} experimental values of $T_{\mathrm{C}}$ are shown together with theoretically calculated and previously reported $T_{\mathrm{C}}$ values for the Fe$_5$SiB$_2$-Fe$_5$PB$_2$ system.
Experimental $T_{\mathrm{C}}$ values show approximately a linear decrease for concentrations $x > 0.1$. The largest experimental $T_{\mathrm{C}}$ is 834 K for Fe$_5$Si$_{0.9}$P$_{0.1}$B$_2$ and the smallest value is 615 K for Fe$_5$PB$_2$. 

$T_\mathrm{C}$ was extracted by drawing the tangent at the inflection point on the $M-T$--curve and determining $T_\mathrm{C}$ from the intercept with the temperature axis. 
This method has given reliable values for other magnetic material systems, e.g.  Gd.\cite{Dankov1998} 
This practice was employed in order to compensate for the Fe$_3$Si impurity phase, a ferromagnetic compound with a $T_\mathrm{C}$ of about 840 K.\cite{shinjo_magnetic_1963,liu2011}
Since $T_\mathrm{C}$ for Fe$_3$Si is higher than for some of the investigated compounds, the temperature dependent magnetization curve will be a superposition of two $M-T$--curves.
Compared to common practice, the uncertainty of this method was estimated to be around $\pm$ 5 K since it changed the extracted $T_\mathrm{C}$ by at most 5 K for the compounds with no visible content of the Fe$_3$Si impurity phase. The calculated Curie temperatures are of similar magnitude as the experimentally measured ones and correctly describe the observation that Fe$_5$PB$_2$ has lower $T_\mathrm{C}$ than Fe$_5$SiB$_2$.

\begin{figure}
	\centering
	\includegraphics[width=0.45\textwidth]{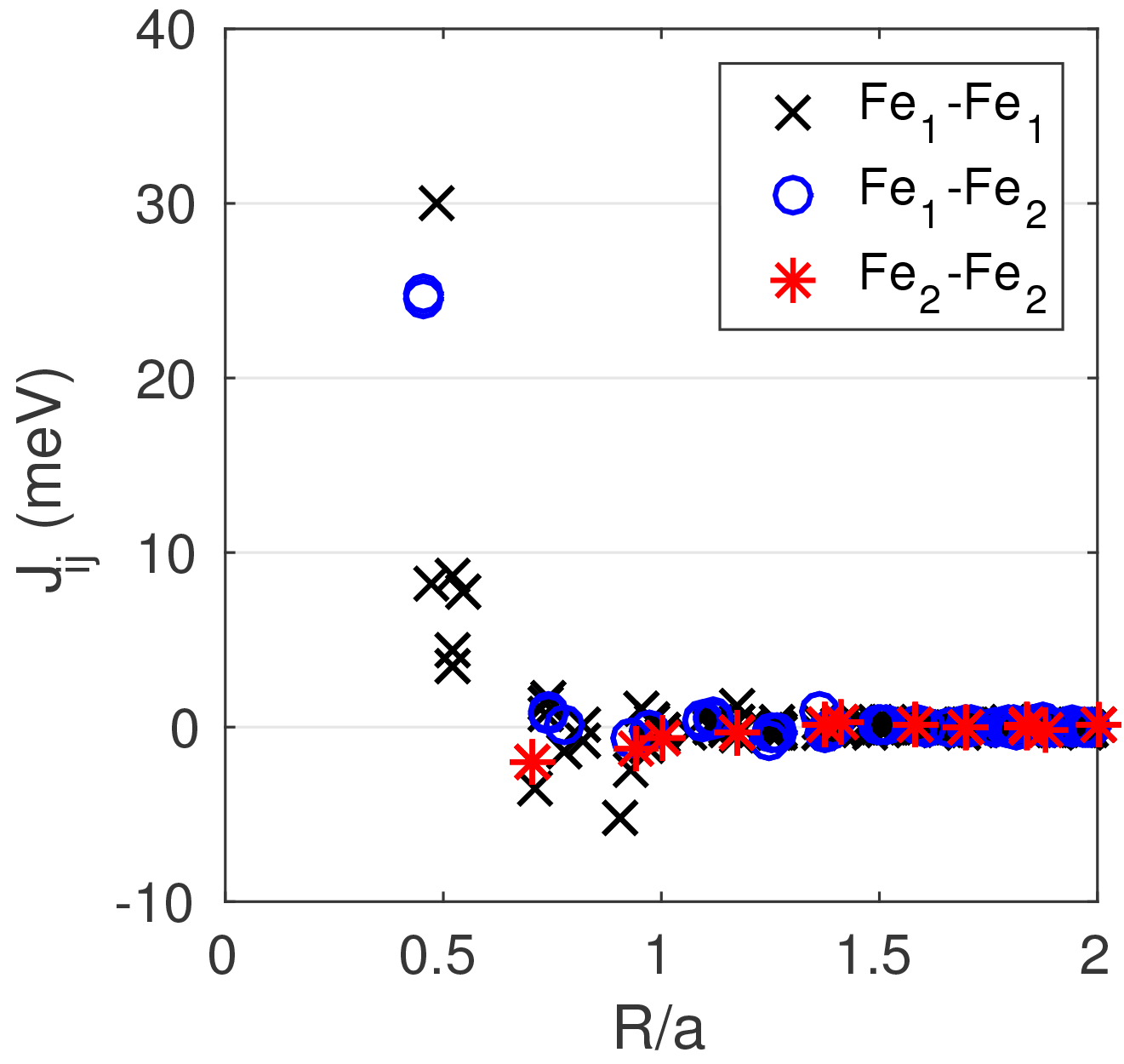}
	\includegraphics[width=0.45\textwidth]{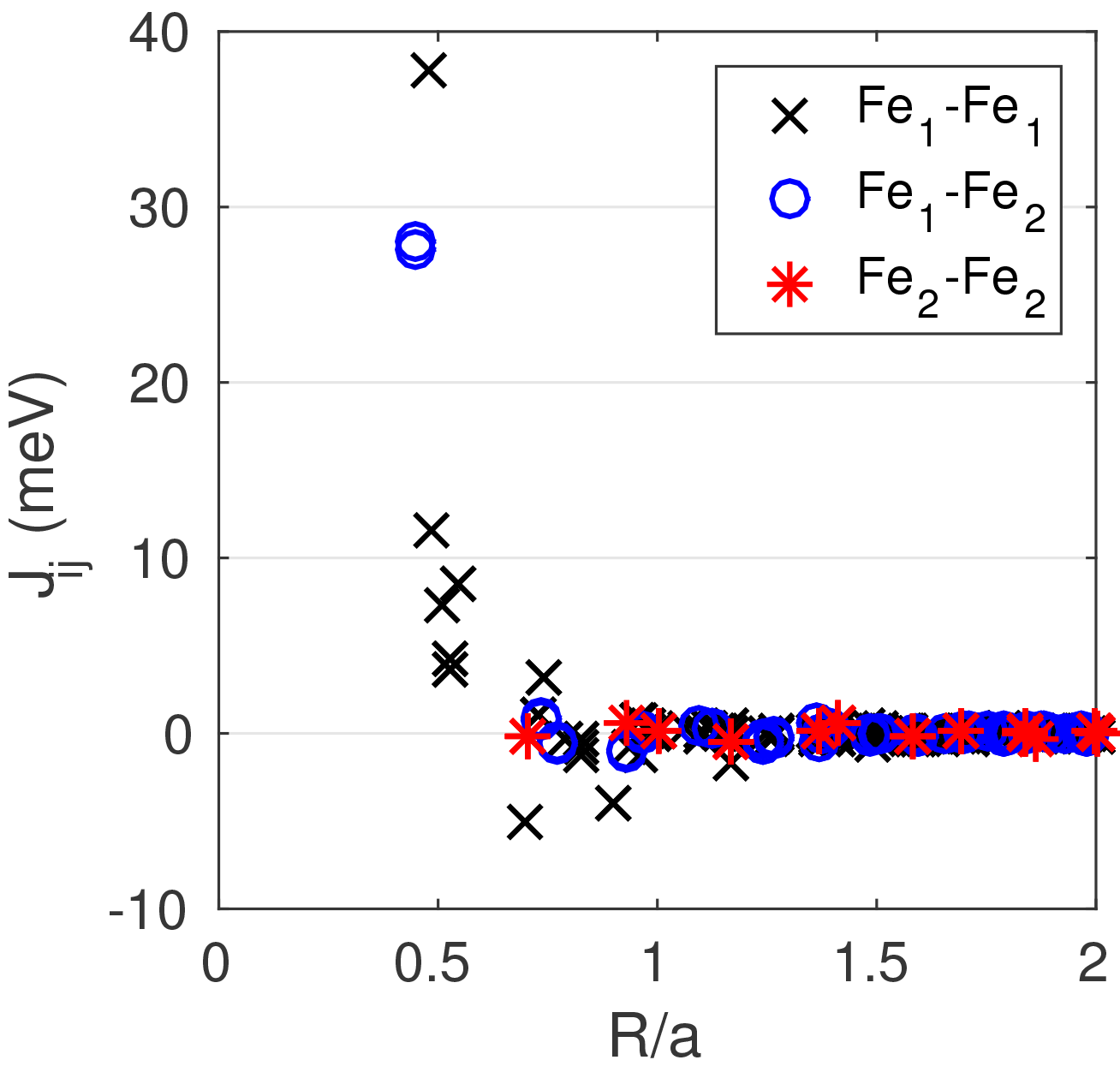}
\caption{Fe-Fe exchange interactions plotted as a function of distance between Fe atoms, for the compounds Fe$_5$PB$_2$ (left) and  Fe$_5$SiB$_2$ (right).}
\label{Fig.Jij}
\end{figure}

Figure~\ref{Fig.Jij} shows the Fe-Fe exchange interactions in the Fe$_5$SiB$_2$ and Fe$_5$PB$_2$ compounds obtained from DFT calculations. The dominating short-range interactions are positive, whereby ferromagnetism is expected to be favored.
Furthermore, several of the dominating interactions can be seen to decrease as P is substituted for Si, which explains the mechanism behind the experimentally observed decrease in $T_\mathrm{C}$.
It is also noticed that all $J^{\text{Fe}_2 - \text{Fe}_2}$ interactions are very weak, and furthermore that there are fewer of these atoms, whereby these should be less important for the magnetic ordering, while Fe$_1$-Fe$_1$ interactions should dominate. This is consistent with the observation that the Fe$_2$ magnetic moments are more strongly reduced with temperature than the Fe$_1$ moments.\cite{Cedervall2016}

Using the exchange coupling parameters presented in Fig.~\ref{Fig.Jij} as input, $T_\mathrm{C}$ has been extracted from classical Monte Carlo (MC) simulations of the Heisenberg Hamiltonian in the UppASD code.\cite{Skubic2008}
System sizes of up to $10^4$ atoms with periodic boundary conditions and exchange interactions up to distances of $3a$, where $a$ is the in-plane lattice parameter, were considered.
Table~\ref{table_Tc} contains the Curie temperatures evaluated from MC simulations as well as those obtained from mean field theory\cite{Anderson196399} and experiments. 
The mean field (MFT) values overestimate the $T_\mathrm{C}$ as is expected, but qualitatively they succeed to describe the higher $T_\text{C}$ in the Si case.
The MC results provide a reasonable agreement with experimental results, both quantitatively and qualitatively.
As pointed out earlier in the text, the decrease in $T_\text{C}$ when substituting P for Si can be understood from the reduction of the short-range $J_\text{ij}$'s observed in Fig.~\ref{Fig.Jij}.
 \begin{table}[h]
\caption{\label{table_Tc} $T_\text{C}$ of Fe$_5$SiB$_2$ and Fe$_5$PB$_2$, calculated from MC simulations and MFT, as well as the experimental values. }

\begin{tabular}{ lrr  }
 \hline \hline
  & Fe$_5$SiB$_2$ & Fe$_5$PB$_2$ \\
\hline 
 $T_\text{C}^{\text{MFT}}$ (K) & 1126 & 990 \\
 $T_\text{C}^{\text{MC}}$ (K) & 700 & 660 \\
 $T_\text{C}^{\text{exp}}$ (K) & 810 & 615 \\
\hline \hline
\end{tabular}
 \end{table} 

For Fe$_5$PB$_2$, the $T_\text{C}$ value of 615 K can be compared to 640 K reported by McGuire \emph{et al.},\cite{McGuire2015} while Lamichhane \emph{et al.} \cite{lamichhane_study_2016} reported a $T_\mathrm{C}$ of 655 K for single crystals. MC calculations in this work result in a slightly higher $T_\text{C}$ of 660 K, while MFT based calculations give 990 K. 
The $T_\text{C}$ data for Fe$_5$SiB$_2$ exhibit larger discrepancies. The experimental $T_\mathrm{C}$ of 810 K found in this study can be compared to the work by McGuire \emph{et al.}\cite{McGuire2015} who reported $T_\mathrm{C}$\textgreater 780 K due to limitations of their setup. The MC based calculation underestimates $T_\mathrm{C}$ whereas the MFT based calculation overestimates the transition temperature compared to the experimental data. 

\begin{figure} 
	\centering
 \includegraphics[width=0.80\textwidth]{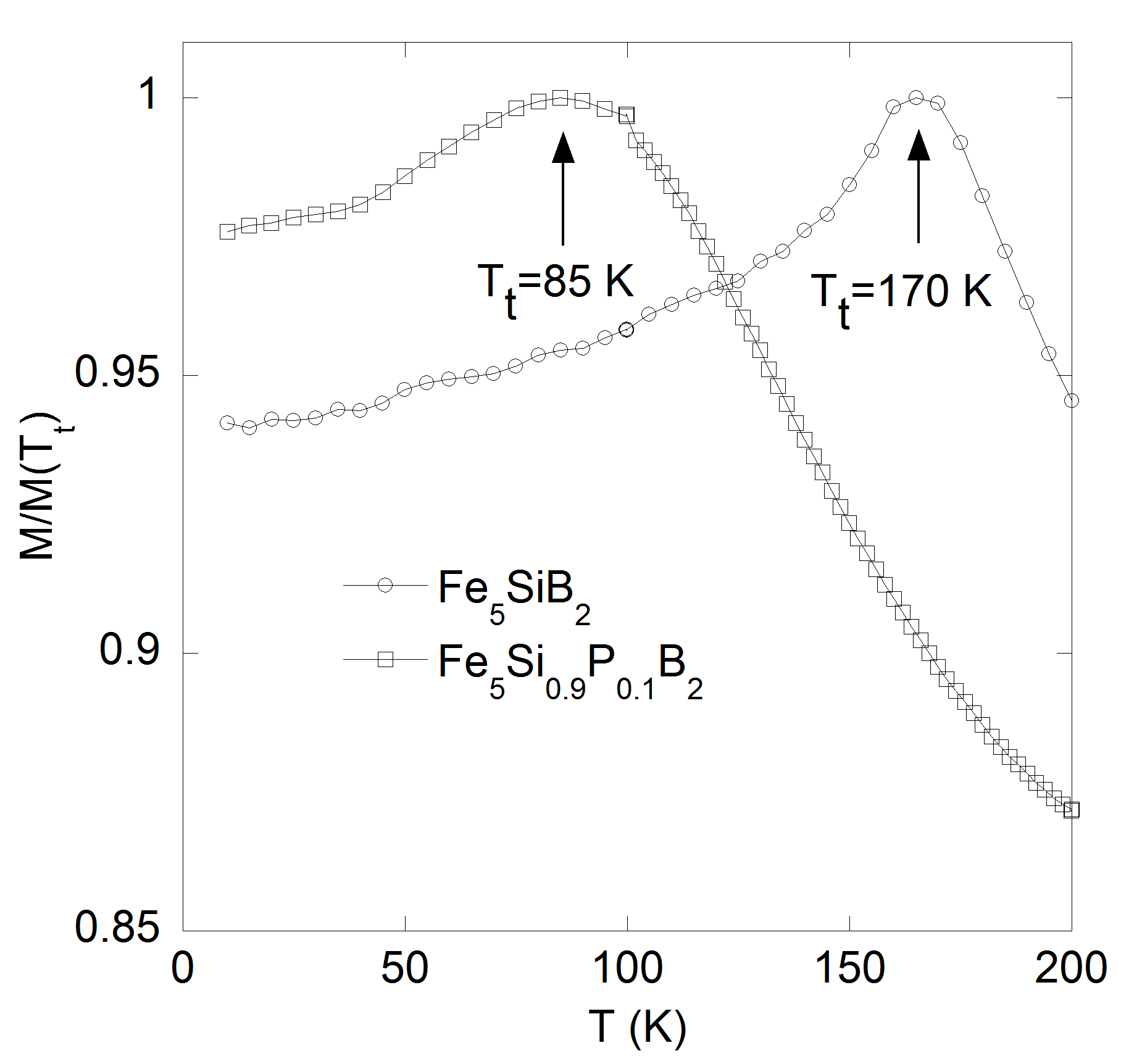}
 \caption{Magnetization, normalized with the magnetization value at the peak temperature $T_t$, vs. temperature for Fe$_5$SiB$_2$ and Fe$_5$Si$_{0.9}$P$_{0.1}$B$_2$.}
 \label{fig:lowt}
\end{figure}

Fig.  \ref{fig:lowt} shows the low-field $M$ vs. $T$ curves for Fe$_5$SiB$_2$ and Fe$_5$Si$_{0.9}$P$_{0.1}$B$_2$.  Both magnetization curves show a maximum at a temperature $T_{t}$; $T_{t}=172$ K and $T_{t}=85$ K for Fe$_5$SiB$_2$ and Fe$_5$Si$_{0.9}$P$_{0.1}$B$_2$, respectively, even though the peak for Fe$_5$Si$_{0.9}$P$_{0.1}$B$_2$ appears more broad. The $M$ vs. $T$ curve for Fe$_5$Si$_{0.8}$P$_{0.2}$B$_2$ exhibits an even broader peak, located in the interval 4 K - 10 K (not shown). In order to observe this peak for the $x=0.2$ sample, the temperature range was extended down to 4 K. The $M$ vs. $T$ curves of the other compounds are featureless within the temperature range covered by the experiment. Previously, using neutron diffraction on Fe$_5$SiB$_2$ a spin--flip transition\cite{Cedervall2016} was revealed where the magnetization on cooling switches from easy--axis to easy--plane. The peaks observed for the three compositions; $x=0$, $x=0.1$ and $x=0.2$, thus give further support for a spin-flip transition in these compounds. The origin of the peak in the low-field magnetization (cf. Fig. \ref{fig:lowt}) is related to the zero-crossing of the anisotropy constant $K_1$ switching the system from an easy--axis to an easy--plane state. At this temperature, the compound would exhibit its largest field induced magnetization. The broadening of the peaks with increasing $x$ may reflect local variations of the MAE or a gradual spin reorientation towards an easy-plane rather than a spin--flip transition at a well defined temperature. However, to fully resolve this issue, further experimental studies are necessary, e.g., using neutron diffraction.

The MAE of a tetragonal system can be described by\cite{coey_magnetism_2010}
\begin{equation}
E = K_1 \sin^2{\theta} + K_2 \sin^4{\theta}, 
\end{equation}
where $K_1$ and $K_2$ are anisotropy constants, and $\theta$ is the angle between magnetization and the easy axis, i.e. the c--axis. Information relating to the magnetocrystalline anisotropy can be derived from the law of approach to saturation\cite{chikazumi_physics_1997} that describes how $M$ varies close to saturation
\begin{equation}
	\frac{M}{M_{\mathrm{S}}} = \left( 1 - \frac{b}{H^2} \right).
\end{equation}
Using the relations obtained by Andreev\cite{andreev_law_1997} the parameter $b$ is given by $b = \frac{4}{15} \left( \frac{ K_{\mathrm{eff}}}{\mu_0 M_{\mathrm{S}}} \right)^2$, which gives $\left| K_{\mathrm{eff}} \right| = \sqrt{ \frac{15b}{4} } \mu_0 M_{\mathrm{S}}$, where the effective value of the anisotropy constant is defined as
\begin{equation}
	K_{\mathrm{eff}}^{2} = K_{\mathrm{1}}^{2}+\frac{16}{7}\left( K_{\mathrm{1}} + \frac{2}{3} K_{\mathrm{2}}\right) K_{\mathrm{2}}.
\end{equation} 

This means that a plot of $\frac{M}{M_{\mathrm{S}}}$ vs. $\frac{1}{H^2}$ results in a straight line  and a numerical value of the regression constant $b$ can be determined which then gives $\left| K_{\mathrm{eff}} \right|$. The selection of regression interval will some have impact on the extracted $\left| K_{\mathrm{eff}} \right|$. In this study, the interval 93--98\% of $M_{\mathrm{S}}$ was used for all regressions. This practice was employed since extending the interval to \textless 93\% introduces deviations from the model, which requires additional terms to describe the behavior of the magnetization curves. Above 98\%, a paramagnetic like term (linear in field) is needed, which is small but without such a term the fitting becomes poor. 
It is estimated that the uncertainty by using other intervals, e.g. 94--97\% and 95--98\%, would give errors no larger than 10\%.  

\begin{figure}[t!]
	\centering
 \includegraphics[width = 0.80\textwidth]{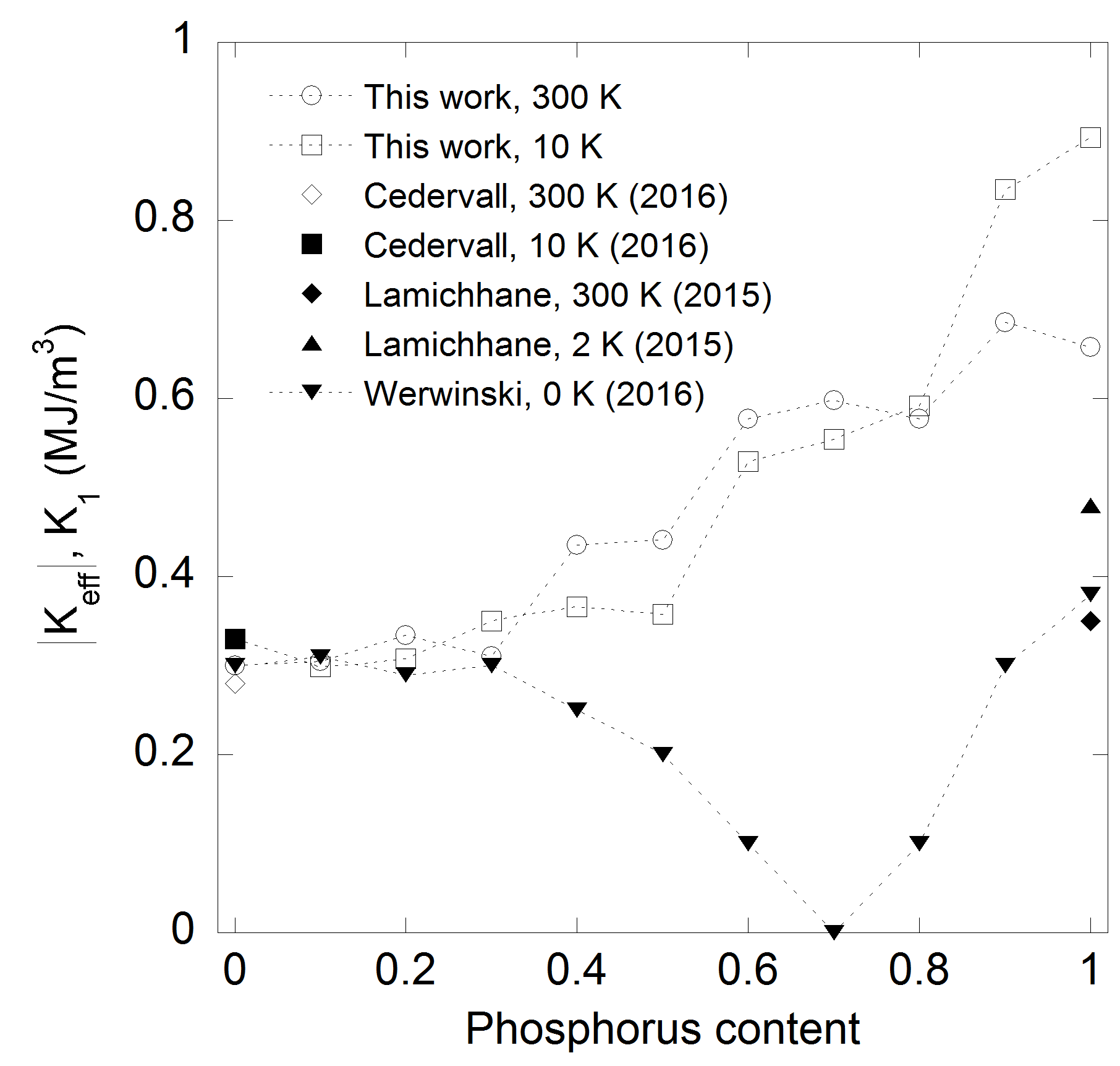}
 \caption{Experimental values for $K_{\mathrm{eff}}$ at 300 K  and 10 K together
 with previously reported values \cite{Cedervall2016}. Note that only the magnitude and not the sign is presented here and that the theoretical work by Werwi\'nski\cite{PhysRevB.93.174412} reports values calculated for zero temperature.}
 \label{fig:keff}
\end{figure}

Fig.  \ref{fig:keff} displays the experimental values of $\left| K_{\mathrm{eff}} \right|$ vs. $x$ for $T=300$ K and $T=10$ K from this study together with values of $K_1$ reported by others. The theoretical values of $K_1$ from Werwi\'nski \emph{et al.},\cite{PhysRevB.93.174412} correspond to zero temperature values. It should be emphasized that $K_{\mathrm{eff}}$ does not generally equal $K_1$. However, $K_2 << K_1$ often holds, in which case $K_{\mathrm{eff}} \approx K_1$. Further, for the comparison with $\left| K_{\mathrm{eff}} \right|$, the calculated $K_1$-values are given as magnitude values, keeping in mind that from theory these values are negative for $x < 0.7$, and for experiment they are negative for $x = 0.2$ and lower and positive for phosphorus rich samples. The negative theoretical values of the MAE is supported by the spin-flip transition observed for Fe$_5$SiB$_2$\cite{Cedervall2016} as well as by the peaks in the $M$ vs. $T$ curves observed for $x = 0, 0.1, 0.2$. It is also worth pointing out that there is no obvious contradiction between the theoretical and experimental results, since the former correspond to zero temperature and the latter to finite temperature results. The peak temperature in the $M$ vs. $T$ curve steeply decreases with increasing $x$ and is below 10 K already for $x = 0.2$, which could explain why the easy--axis to easy--plane transition was not observed in experiments for phosphorus concentrations $0.2 < x < 0.7$.

Comparing with the data obtained for single crystals by Lamichhane \emph{et al.}\cite{lamichhane_study_2016}, the value obtained here for $\left| K_{\mathrm{eff}} \right|$ is 70\% larger than the $K_1$ value obtained for single crystalline Fe$_5$PB$_2$. The agreement for low P concentrations is much better indicating that the data obtained using the law of approach to saturation gives quantitatively correct values for low P concentrations and low temperatures as seen in Fig. \ref{fig:keff}. 

One motivation for this study was to investigate if the system for some range of phosphorus concentration would exhibit properties making the materials suitable candidates as rare-earth free permanent magnets. The aim to design new permanent magnets with sufficiently large MAE is a highly non-trivial task, from a synthesis and characterization point of view, but also from theory. Some guiding principles have however been identified, for how to find itinerant electron magnets with large MAE. As argued in Refs. [\onlinecite{Costa2012,Edstrom1033475}], one should combine heavy elements with a large spin-orbit coupling constant, with 3d elements that provide a large exchange splitting and hence large saturation moments and ordering temperature. In addition, a careful tuning of the bandstructure should be made, in such a way that weakly dispersive electron states are found on each side of the Fermi level, with a small energy separation. Although this is something that comes out of theoretical considerations of relativistic energy band structures, it may not directly guide an experimental realization of new magnets. It offers however, a possibility for theoretical and experimental efforts to join forces, in order to reach goals that otherwise might not have been possible. The present study is an example of such a joint effort, and although the materials identified here are not suitable as permanent magnets, the results of this investigation are still encouraging since it shows that theory and experiment agree on most of the relevant magnetic properties.

\section{Conclusions}
The structural and magnetic properties of the Fe$_5$SiB$_2$-Fe$_5$PB$_2$ system have been studied experimentally as well as by performing DFT calculations. The system crystallizes in the $I$4/$mcm$ space group, the $a$ lattice parameter decreases linearly from $5.5541(1)~\text{\AA{} }$ to $5.4923(1)~\text{\AA{}}$ while the $c$ parameter increases slightly with increasing phosphorous substitution. All compounds are ferromagnetic with a maximum $T_\mathrm{C}$ of 837 K for Fe$_5$Si$_{0.9}$P$_{0.1}$B$_2$, in good agreement with the results from MC calculations. The saturation magnetization has been found to decrease from 1.03 MA/m to 0.87 MA/m at 300 K as Si is substituted for P. Substitution of Si by P does not change the crystal structure, but the volume of the unit cell changes almost linearly with changing P concentration. The phosphorus rich compounds exhibit higher MAE according to the method of law of approach to saturation, which only gives the magnitude and not the sign of the MAE. Results from zero temperature VCA calculations performed by Werwi\'nski \emph{et al.},\cite{PhysRevB.93.174412} indicate that the MAE should cross zero at $x \approx 0.7$, implying that the system will be in an easy--plane state for smaller $x$-values. The negative theoretical values of the MAE for phosphorus poor samples are supported by the spin-flip transition observed for Fe$_5$SiB$_2$\cite{Cedervall2016} as well as by the peaks in the $M$ vs. $T$ curves observed for $x = 0, 0.1, 0.2$.
The peak temperature in the $M$ vs. $T$ curve steeply decreases with increasing $x$ and it is below 10 K already for $x = 0.2$, which could explain why the easy--axis to easy--plane transition was not observed in experiments for phosphorus concentrations $0.2 < x < 0.7$. A quantitative comparison between experimental and theoretical results is complicated by the fact that the theoretical results correspond to zero temperature results, while the experimental results have been obtained at finite temperature. However, the correct easy axis/easy plane is obtained in the theoretical calculations of phosphorus rich and phosphorus poor samples, with an anisotropy energy that is qualitatively in agreement with experimental observations. Comparing the present MAE experimental results with results obtained for single crystals of Fe$_5$PB$_2$ by Lamichhane \emph{et al.} \cite{lamichhane_study_2016}, reveal good agreement for $x=0$, while the MAE obtained here for $x=1$ is considerably larger.

\section*{Acknowledgment}
This work was supported by Swedish Research Council and G. Gustafsson's foundation.  O. Eriksson acknowledges support from the KAW Foundation (projects No. 2013.0020 and No. 2012.0031) and EU Horizon 2020 program NOVAMAG. Calculations performed from an allocation of SNIC. M. Werwi\'nski acknowledges the financial support from the Foundation of Polish Science grant HOMING. The HOMING programme is cofinanced by the European Union under the European Regional Development Fund.

\bibliographystyle{apsrev4-1}

\end{document}